\documentclass{PoS}

\usepackage{amsmath}

\newcommand\cO{\mathcal{O}} 
\newcommand{\third}{\mbox{\small $\frac{1}{3}$}}
\newcommand{\Dd}[1]{\overset{\leftrightarrow}{D}_{#1}}

\newcommand{\MS}{{\overline{\mathrm{MS}}}}
\newcommand{\RI}{{\mathrm{RI}^\prime - \mathrm {MOM}}}

\newcommand{\momt}{\widetilde{\mathrm {MOM}}\mathrm{gg}}

\title{Renormalisation of composite operators in lattice QCD:
perturbative versus nonperturbative}

\ShortTitle{Renormalisation of composite operators in lattice QCD}

\author{\speaker{M.~G\"ockeler}$^a$, R.~Horsley$^b$, Y.~Nakamura$^a$,
H.~Perlt$^c$, D.~Pleiter$^d$, P.E.L.~Rakow$^e$, A.~Sch\"afer$^a$,
G.~Schierholz$^{af}$, A.~Schiller$^c$, H.~St\"uben$^g$ and J.M.~Zanotti$^b$ \\
\llap{$^a$} Institut f\"ur Theoretische Physik,
Universit\"at Regensburg, 93040 Regensburg, Germany \\
\llap{$^b$} School of Physics and Astronomy,
University of Edinburgh, Edinburgh EH9 3JZ, UK \\
\llap{$^c$} Institut f\"ur Theoretische Physik,
Universit\"at Leipzig, 04109 Leipzig, Germany \\
\llap{$^d$} Deutsches Elektronen-Synchrotron DESY 
and John von Neumann-Institut f\"ur Computing NIC, 15738  Zeuthen, Germany \\
\llap{$^e$} Theoretical Physics Division,
Department of Mathematical Sciences, University of Liverpool, 
Liverpool L69 3BX, UK \\
\llap{$^f$} Deutsches Elektronen-Synchrotron DESY, 22603 Hamburg, Germany \\
\llap{$^g$} Konrad-Zuse-Zentrum f\"ur Informationstechnik Berlin,
14195 Berlin, Germany \\

E-mail: \email{meinulf.goeckeler@physik.uni-regensburg.de} }

\author{QCDSF/UKQCD Collaboration}

\abstract{
The perturbative and nonperturbative renormalisation of
quark-antiquark operators in lattice QCD with two flavours of
clover fermions is investigated within the research programme of
the QCDSF collaboration. Operators with up to three derivatives
are considered. The nonperturbative results based on the RI-MOM
scheme are compared with estimates from one- and two-loop lattice
perturbation theory.

\begin{flushright} Preprint DESY 10-161, Edinburgh 2010/23, LU-ITP 2010/005
\end{flushright}
}

\FullConference{The XXVIII International Symposium on Lattice Field Theory, Lattice2010\\
		June 14-19, 2010\\
		Villasimius, Italy}

\begin{document}

\section{Introduction}
If one uses a lattice regulator to define QCD, renormalisation means that 
bare parameters, i.e., the coupling constant and the quark masses, acquire
a dependence on the lattice spacing $a$ such that physical observables like
hadron masses have a finite continuum limit $a \to 0$. This should then 
guarantee that all physical quantities have a well-defined continuum limit.

However, it is not always possible to calculate the 
observables of interest directly on the lattice. Quite often one can only 
evaluate certain matrix elements of composite operators from which the
desired physical quantity is derived in a second step. In general, it 
is then necessary to introduce $a$-dependent renormalisation factors $Z$ 
for these operators in order to get a finite continuum limit. 
For example, moments of structure functions in deep-inelastic 
scattering are physical observables, but they can only be calculated
as products of a (perturbatively calculated) Wilson coefficient and 
a hadronic matrix element of a local composite operator, which is a
long-distance quantity to be computed on the lattice. 

It is therefore important to investigate the renormalisation of
composite operators in lattice QCD, and the present contribution 
describes some of the efforts the QCDSF collaboration has undertaken
in this field. For the details we refer to Ref.~\cite{npr}.

\section{How to evaluate renormalisation factors on the lattice}

In principle it is possible to calculate the $Z$ factors of 
multiplicatively renormalisable operators by lattice 
perturbation theory (for a review see Ref.~\cite{stefano}). 
However, perturbation theory on the lattice is computationally 
much more complex than in the continuum and 
therefore the calculations rarely extend beyond one-loop order 
(see, however, Refs.~\cite{mason,panagopoulos1,panagopoulos2,stoch}). 
Moreover, lattice perturbation theory usually converges rather 
slowly so that the accuracy of perturbative renormalisation factors 
is limited, even if some improvement scheme is applied. 

Therefore nonperturbative approaches have been developed, in 
particular methods based on the Schr\"odinger functional (for
reviews see Ref.~\cite{sommer}) and the 
RI-MOM scheme~\cite{marti}. It is the latter approach in the
slightly modified form of the $\RI$ scheme that was
adopted to produce the nonperturbative results presented below,
where we consider only composite operators constructed from two
quark fields, Dirac matrices and a few covariant derivatives in 
between. 

In the $\RI$ scheme the basic objects are quark two-point functions 
with an insertion of the operator under consideration at momentum zero. 
A suitable renormalisation
condition is imposed, which does not rely on a particular regularisation 
so that it can be applied on the lattice as well as in the continuum. 
The latter property entails the possibility to use continuum
perturbation theory in order to calculate conversion 
factors leading from the operators renormalised in the $\RI$ scheme
to operators renormalised in, e.g., the $\MS$ scheme.

For a multiplicatively renormalisable operator we express the
operator renormalised in the scheme $\mathcal S$ at the renormalisation
scale $\mu$ as $Z_{\mathrm {bare}}^{\mathcal S} (\mu) \cO (a)$, 
where $\cO (a)$ denotes the bare operator on the lattice and the $a$
dependence of the renormalisation factor has been suppressed. 
For scales satisfying
$\Lambda^2_{\mathrm {QCD}} \ll \mu^2 \ll 1/a^2$ the quantity
\begin{equation} \label{defrgi}
Z^{\mathrm {RGI}} 
= \left( 2 \beta_0 \frac{g^{\mathcal S} (\mu)^2}{16 \pi^2}
\right)^{-\gamma_0/(2 \beta_0)} 
 \exp \left \{ \int_0^{g^{\mathcal S} (\mu)} \! \mathrm d g \, 
\left( \frac{\gamma^{\mathcal S} (g)}{\beta^{\mathcal S} (g)} 
                                  + \frac{\gamma_0}{\beta_0 g} 
\right) \right \} Z_{\mathrm {bare}}^{\mathcal S} (\mu)
 = \Delta Z^{\mathcal S} (\mu) Z_{\mathrm {bare}}^{\mathcal S} (\mu)
\end{equation}
is independent of the scale and the scheme. Here 
$g^{\mathcal S} (\mu)$ denotes the renormalised coupling in the 
scheme $\mathcal S$, while $\gamma^{\mathcal S}$ and $\beta^{\mathcal S}$
are the anomalous dimension and the beta function, respectively, whose
one-loop coefficients are $\gamma_0$ and $\beta_0$. 
In the simulations we compute $Z_{\mathrm {bare}}^{\RI}$ and 
evaluate
\begin{equation} \label{compzrgi}
Z^{\mathrm {RGI}} 
 = \Delta Z^{\mathcal S} (\mu) Z_{\RI}^{\mathcal S} (\mu)
Z_{\mathrm {bare}}^{\RI} (\mu) \,.
\end{equation}
This expression would be identical to (\ref{defrgi}) if we knew 
$\Delta Z^{\mathcal S}$ and the conversion factor 
$Z_{\RI}^{\mathcal S}$ exactly. However, 
these quantities are calculated in continuum perturbation theory, so
they come with certain truncation errors and the result will
depend on the choice of the intermediate scheme $\mathcal S$.
What will also matter is the expansion parameter. It turns out to
be advantageous not to use $g^{\MS}$ but the coupling $g^{\momt}$
as defined in Ref.~\cite{cheret2}.

In bare lattice perturbation theory one-loop results for renormalisation
factors are of the form 
\begin{equation}
Z_{\mathrm {bare}}^{\mathcal S} (\mu)_{\mathrm {pert}} 
= 1 - \frac{g^2}{16 \pi^2}
\left( \gamma_0 \ln (a \mu) + \Delta \right) + O(g^4) \,. 
\end{equation}
As this expansion in the bare lattice coupling $g$ is often poorly 
convergent, tadpole-improved perturbation theory has been invented,
where for an operator with $n_D$ covariant derivatives one has
\begin{equation}
Z_{\mathrm {bare}}^{\mathcal S} (\mu)_{\mathrm {ti}} 
= u_0^{1-n_D} \left[1 - \frac{g_\Box^2}{16 \pi^2}
\left( \gamma_0 \ln (a \mu) + \Delta + 
  (n_D - 1)\frac{4}{3} \pi^2 \right) + O(g^4) \right] \,.
\end{equation}
Here the fourth root of the average plaquette 
$u_0 = \langle \third \mathrm {tr} U_\Box \rangle ^{1/4}$ is taken
from the simulations and the expansion parameter is the boosted coupling
$g_\Box^2 = g^2/u_0^4$. Combining renormalisation group improvement 
with tadpole improvement one arrives at ``tadpole-improved 
renormalisation-group-improved boosted perturbation theory'' or
TRB perturbation theory for short~\cite{npr}.

In the above expressions for $Z$ derived from lattice perturbation theory 
lattice artefacts have been neglected and only the logarithmic $a$ 
dependence has been kept. This is only justified if $a^2 \mu^2 \ll 1$.
Unfortunately, this condition is not always fulfilled in our simulations.
On the other hand, it is straightforward (though increasingly involved
for more complicated operators) to do calculations in one-loop lattice 
perturbation theory with arbitrary values of $a^2 \mu^2$. One can then
use the difference between the $Z$s with and without lattice artefacts
to correct the nonperturbative simulation results for discretisation errors
of $O(g^2)$. 

\section{The simulations}

We use gauge field configurations generated by the QCDSF-UKQCD 
collaborations with two degenerate flavours of clover fermions and 
the Wilson plaquette action for the gauge field. Configurations for
4 values of $\beta$ are at our disposal, $\beta = 5.20$, 5.25, 5.29, 5.40,
corresponding to lattice spacings $a \approx 0.086$, 0.079, 0.075, 0.067
fm. To set the scale for the lattice spacing we have taken the value
$r_0 = 0.467 \, \mathrm {fm}$ for the Sommer parameter, and we 
use $r_0 \Lambda_{\MS} = 0.617$~\cite{lambda} 
when the $\Lambda$ parameter of QCD is needed. For each $\beta$ 
we have between three and five sea quark masses so that a chiral 
extrapolation to vanishing quark mass is possible. 

As mentioned above, we have to compute quark two-point functions with 
an insertion of the operator under consideration, i.e., three-point 
functions. We restrict ourselves to flavour-nonsinglet operators so that
only quark-line connected contributions must be evaluated. This is 
conveniently done with the help of momentum sources introduced in 
Ref.~\cite{reno}. 

\section{Extracting the renormalisation factors}

\begin{figure}
\begin{center}
\includegraphics[width=.49\textwidth]{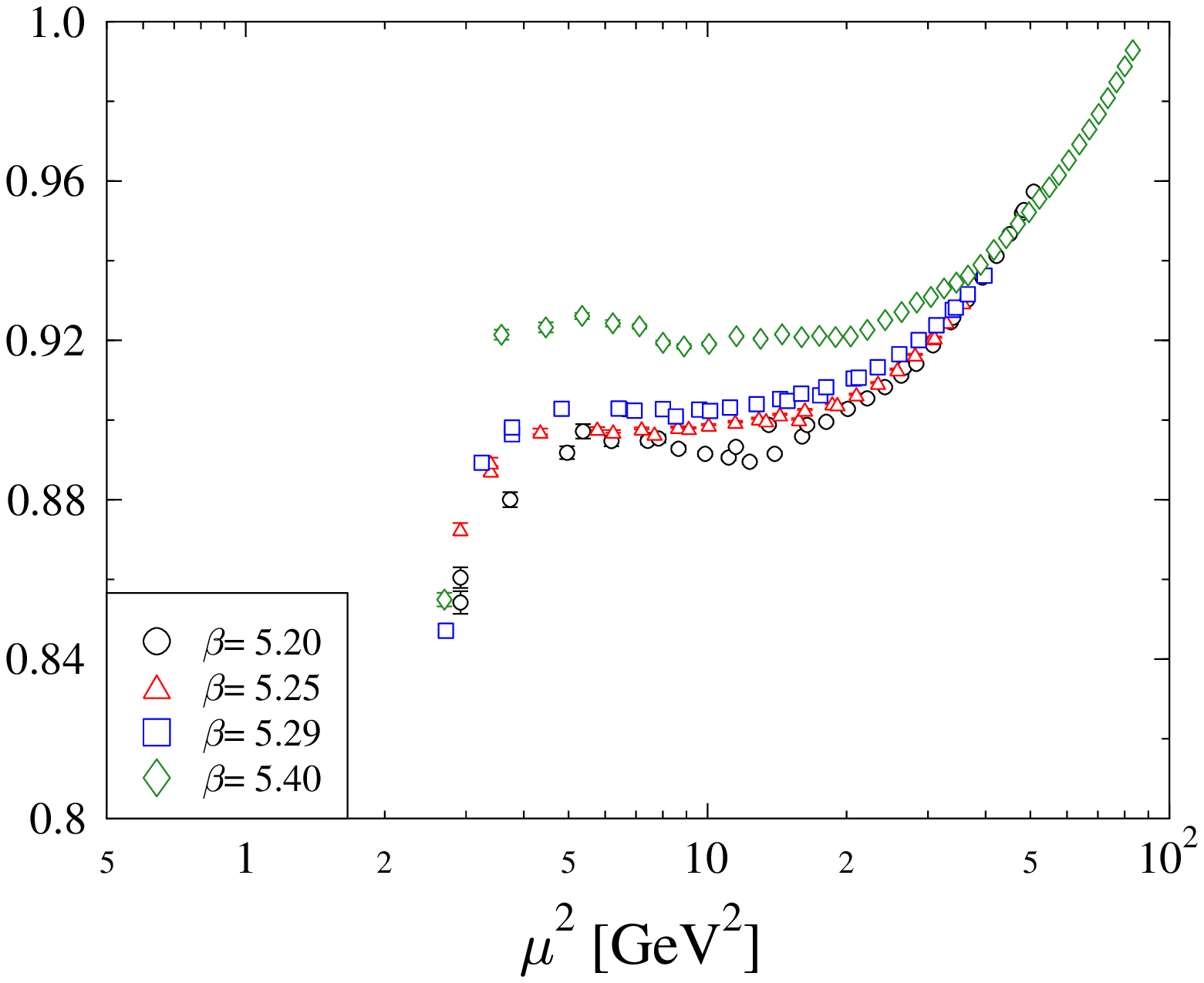}
\includegraphics[width=.49\textwidth]{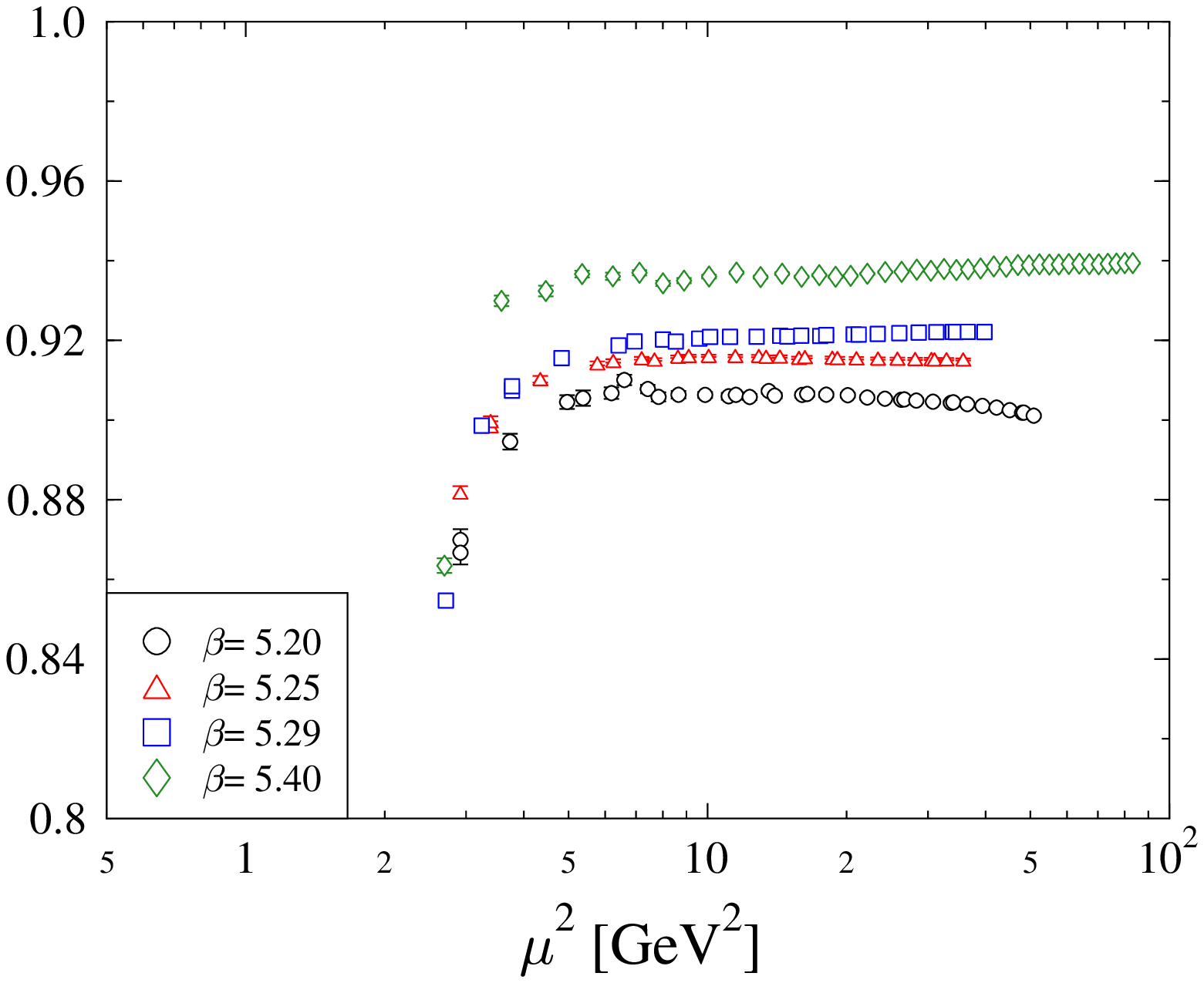}
\end{center}
\caption{$Z^{\mathrm {RGI}}$ for the tensor current 
$\bar{\psi} \sigma_{\mu \nu} \psi$ before (left panel) and after (right 
panel) the perturbative subtraction of lattice artefacts.}
\label{fig.rgits}
\end{figure}

The simplest procedure for obtaining a value of
$Z^{\mathrm {RGI}}$ would be to plot the right-hand side of
Eq.~(\ref{compzrgi}), i.e., 
$\Delta Z^{\mathcal S} (\mu) Z_{\RI}^{\mathcal S} (\mu)
Z_{\mathrm {bare}}^{\RI} (\mu)$ versus $\mu$ and 
to fit a constant to these data in an interval of $\mu$ where they 
form a plateau. Examples of such plots before and after the 
perturbative subtraction of lattice artefacts are shown in 
Fig.~\ref{fig.rgits}. Equivalently one could fit the values obtained for
$Z_{\mathrm {bare}}^{\mathcal S} (\mu)
= Z_{\RI}^{\mathcal S} (\mu) Z_{\mathrm {bare}}^{\RI} (\mu)$
in the plateau region by
$\Delta Z ^{\mathcal S} (\mu)^{-1} Z^{\mathrm {RGI}}$
(with $Z^{\mathrm {RGI}}$ as fit parameter). 

\begin{figure}
\begin{center}
\includegraphics[width=.49\textwidth]{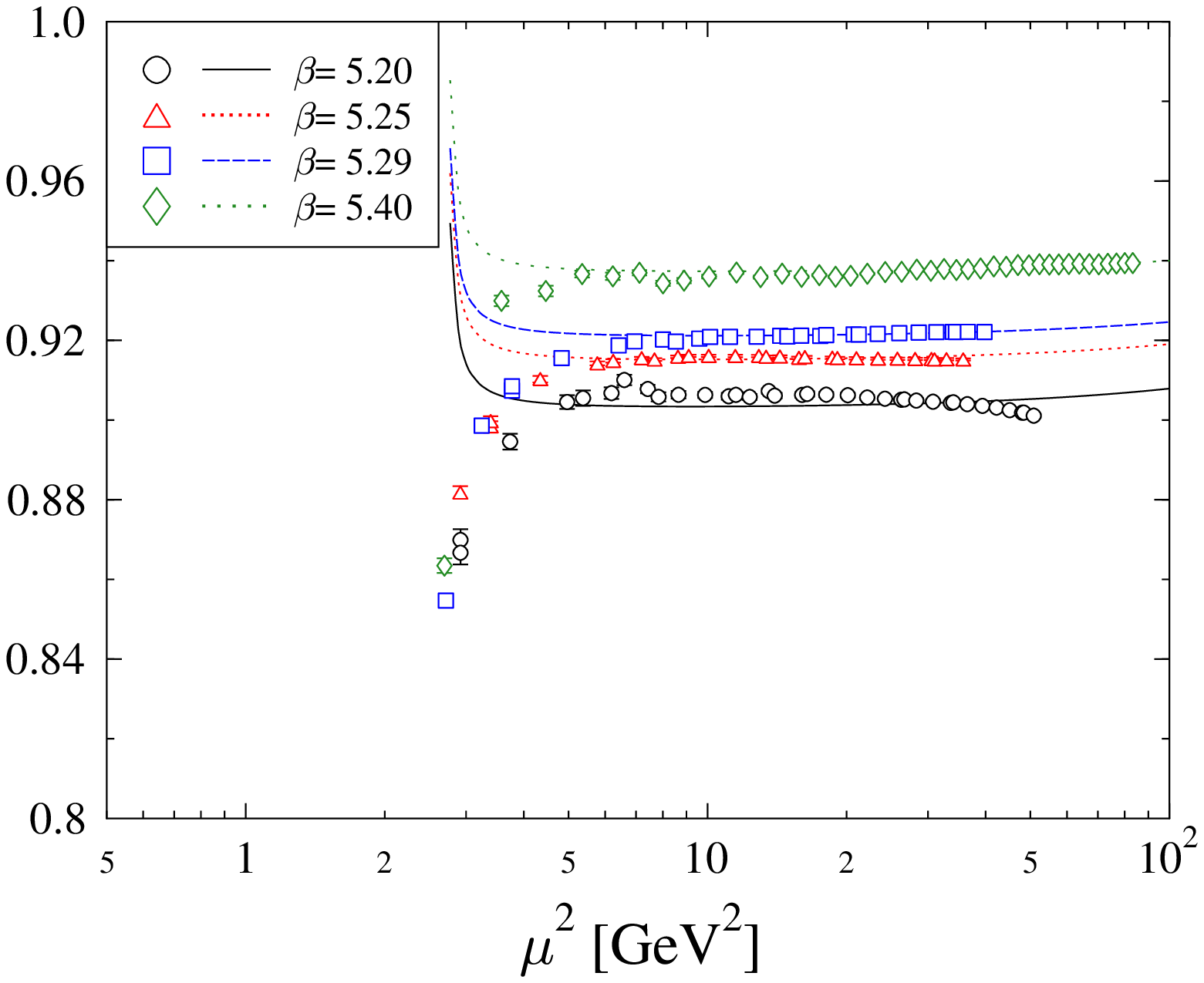}
\includegraphics[width=.49\textwidth]{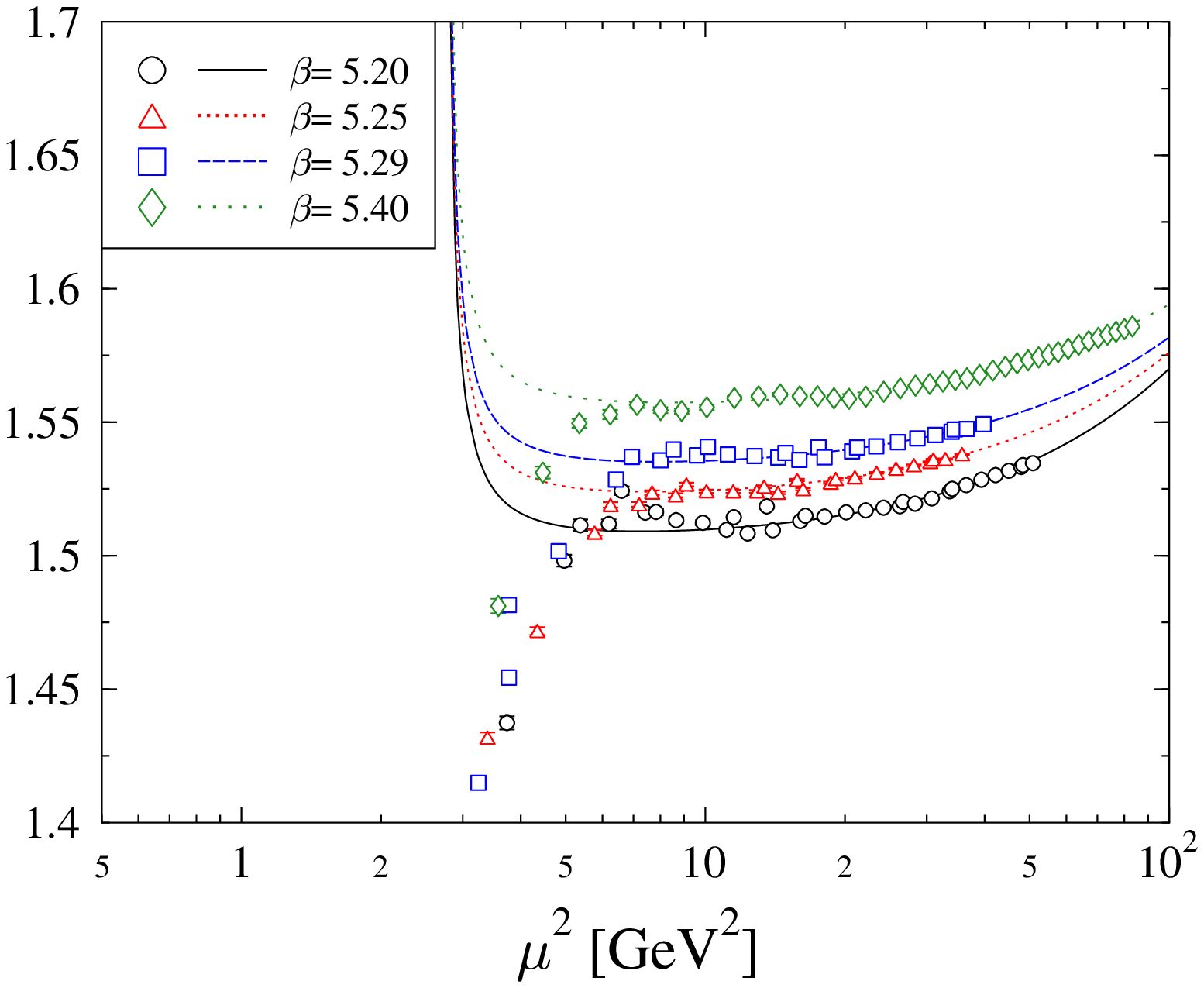}
\end{center}
\caption{$Z^{\mathrm {RGI}}$ (perturbatively subtracted)
for the tensor current (left panel) and 
$\bar{\psi}\gamma_{\{ 1} \protect \Dd{4 \}}\psi$ (right panel) as a 
function of the renormalisation scale. Also shown are the fit curves 
used for the determination of $Z^{\mathrm {RGI}}$.}
\label{fig.fitexamples}
\end{figure}

However, there are two effects that jeopardize the reliability of 
this approach: lattice artefacts, which show up at large values 
of $\mu$ and truncation errors of the perturbative expansions 
in $\Delta Z ^{\mathcal S}$ and $Z_{\RI}^{\mathcal S}$, which 
become noticeable at small values of $\mu$.
They might even conspire to produce a fake plateau. Therefore we 
have tried to incorporate higher terms in the perturbative series
treating the corresponding coefficients as additional fit parameters.
Similarly we have attempted to correct for discretisation effects by
including a simple ansatz for lattice artefacts. Again, the parameters
in this ansatz have to be fitted. 
We fit the data for all
four $\beta$ values simultaneously and only the quantities
$Z^{\mathrm {RGI}}$, our final results, depend on $\beta$,
the other parameters do not. Two examples of such fits are shown in
Fig.~\ref{fig.fitexamples}. For further details concerning the fit 
procedure we must refer to Ref.~\cite{npr}. The results determined
by this fit procedure will be called fit results in the following.

Unfortunately, these fits work only for the perturbatively subtracted 
data. However, for technical reasons, we have applied our subtraction 
only for operators with at most one covariant derivative. Hence we 
must apply a different procedure for operators with more than one 
derivative: We read off $Z^{\mathrm {RGI}}$ at 
$\mu^2 = 20 \, \mbox{GeV}^2$ and take as the error the maximum of 
the differences with the results at $\mu^2 = 10 \, \mbox{GeV}^2$ and 
$\mu^2 = 30 \, \mbox{GeV}^2$. These results will be called interpolation 
results in the following. Obviously, this method can also be
applied to subtracted data.
The errors assigned by the fits appear to be seriously underestimated
since they are mainly determined by the statistical uncertainties.
Therefore it seems to be more reasonable to finally use only the errors
from the interpolation method as these take into account also
some of the systematic effects.

\section{Results}

Due to lack of space we have to restrict the presentation of our 
results to operators without derivatives, i.e.\ the ``currents''
\begin{equation}
\cO^S  =  \bar{u} d \,, \;
\cO^P =  \bar{u} \gamma_5 d \,, \;
\cO^V_\mu  = \bar{u} \gamma_\mu d \,, \;
\cO^A_\mu = \bar{u} \gamma_\mu \gamma_5 d \,, \;
\cO^T_{\mu \nu}  =   \bar{u} \sigma_{\mu \nu} d \,,
\end{equation}
and the quark wave function renormalisation constant $Z_q$.

\begin{figure}
\vspace*{-1.0cm}
\begin{center}
\includegraphics[width=.8\textwidth]{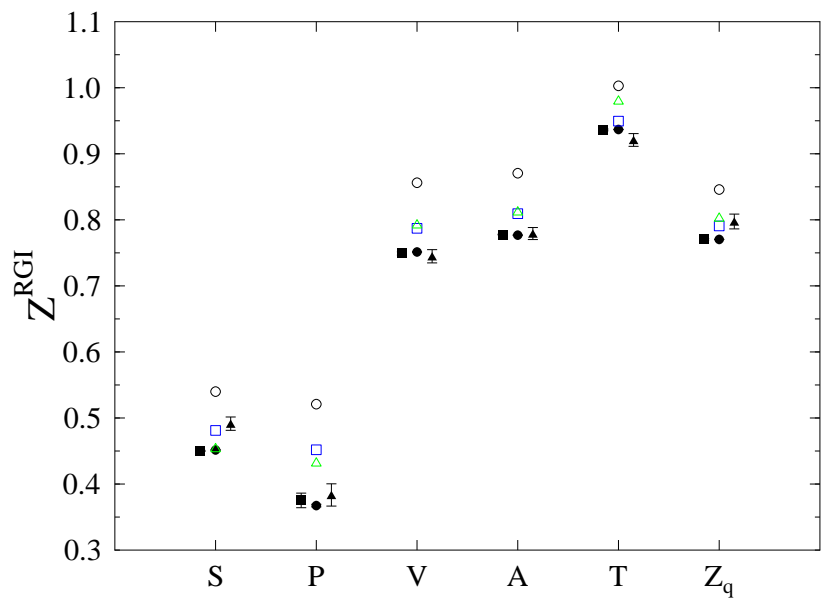}
\end{center}
\vspace*{-1.0cm}
\caption{Results for operators without derivatives at $\beta = 5.40$.
The filled symbols correspond to our fit results (circles), interpolation 
results based on subtracted (squares) and unsubtracted (triangles) data. 
The open symbols represent estimates from bare perturbation theory (circles),
tadpole-improved perturbation theory (squares) and TRB perturbation theory
(triangles) based on one-loop calculations. }
\label{fig.res0}
\end{figure}

In Fig.~\ref{fig.res0} we plot the results at $\beta = 5.40$
extracted from the perturbatively subtracted data,
both by interpolation and by means of the fit procedure, and
the interpolation results based on the unsubtracted numbers
as well as one-loop perturbative estimates.
Ideally the nonperturbative results should agree within the errors. 
In reality, this is not always true. Note, however, that
the errors of the fit results only account for the (rather small)
statistical uncertainties of the raw data while the errors of
the interpolation results are dominated by systematic effects.
The one-loop perturbative estimates are larger than the nonperturbative
values, but tadpole improvement works.
TRB perturbation theory, on the other hand, leads to
further improvement only in a few cases, for some operators it is even
worse than ordinary tadpole-improved perturbation theory.

\begin{figure}
\vspace*{-1.0cm}
\begin{center}
\includegraphics[width=.8\textwidth]{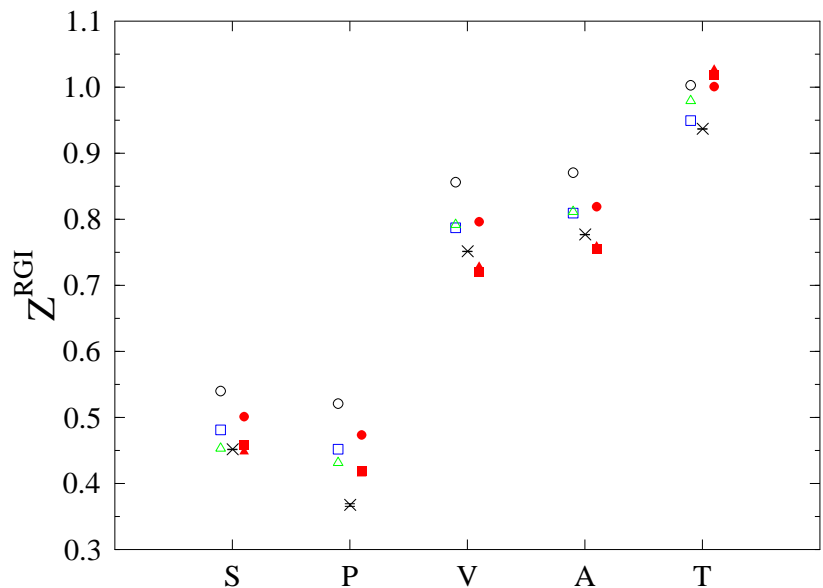}
\end{center}
\vspace*{-1.0cm}
\caption{Results for operators without derivatives at $\beta = 5.40$
compared with one- and two-loop lattice perturbation theory.
The crosses correspond to our nonperturbative results obtained
by fits of the subtracted data. The open symbols represent estimates
from bare perturbation theory (circles), tadpole-improved perturbation
theory (squares) and TRB perturbation theory (triangles) in the
one-loop approximation. The corresponding estimates based on two-loop
calculations are shown by the filled symbols.}
\label{fig.res2loops}
\end{figure}

In Fig.~\ref{fig.res2loops} we compare our fit results
for the operators without derivatives with the one-loop and 
two-loop perturbative estimates, again for $\beta = 5.40$. The numbers from
bare lattice perturbation theory, represented by circles in the
figure, exhibit the expected behavior: The two-loop results come
closer to the nonperturbative numbers than the one-loop estimates,
though only slightly in the case of the tensor current.
Except for the tensor current, tadpole improvement
works also in the two-loop approximation moving the perturbative
values closer to the nonperturbative numbers.
However, the results from TRB perturbation theory, shown by triangles,
do not differ much from the values found by tadpole improved
two-loop perturbation theory. 

\section{Concluding remarks}

The $\RI$ scheme has been established as a method for nonperturbative 
renormalisation that can (relatively) easily be implemented for
arbitrary lattice fermions. Momentum sources allow us to deal with
all operators in a single simulation and at the same time to achieve
small statistical errors, but the required computer time is proportional 
to the number of momenta considered. One of the largest sources of
systematic uncertainties are discretisation effects: Here the perturbative
subtraction of lattice artefacts has proved very helpful. Continuum
perturbation theory is needed for the conversion to the $\MS$ scheme.
Obviously, one should use as many loops as are available, but additional
improvements are possible through the careful choice of intermediate
schemes and expansion parameters. Still, it seems that the available 
perturbative results cannot describe the scale dependence below the 
(surprisingly large) value of about $5 \, \mbox{GeV}^2$, as seen, e.g.,
in Fig.~\ref{fig.rgits}. Remarkably enough, there are now renormalisation
factors calculated in two-loop lattice perturbation theory. 
Unfortunately, it is difficult to predict their accuracy without
comparing with nonperturbative results, but (at least in most cases) 
improvement seems to work.

\section*{Acknowledgements}
The numerical calculations have been performed on the
apeNEXT and APEmille computers at NIC/DESY (Zeuthen).
This work has been supported in part by the EU Integrated Infrastructure
Initiative HadronPhysics2 and by the DFG (SFB/TR55
``Hadron Physics from Lattice QCD'').

\end{document}